\documentclass[twocolumn,showpacs,preprintnumbers,amsmath,amssymb,prl,showkeys]{revtex4}

\usepackage{graphicx}

\begin{document}

\title{%
Exact Solutions of Domain Wall and Spiral Ground States in Hubbard
Models
}

\author{Makoto Homma}

\author{Chigak Itoi}

\affiliation{%
Department of Physics, Nihon University, Kanda, Surugadai, Chiyoda,
Tokyo, Japan%
}

\begin{abstract}
We construct a set of exact ground states with a localized ferromagnetic
domain wall and an extended spiral structure in a 
deformed flat-band Hubbard model. In the case of quarter filling, we
show the uniqueness of the ground state with a fixed magnetization. We
discuss a more realistic situation 
given by a band-bending perturbation, which can stabilize these curious
structures. We study a conduction electron scattered by the
domain wall and the spiral spins.
\end{abstract}


\maketitle


Domain wall and spiral structures in ferromagnetic systems are
interesting structures. Generally a domain wall is localized stably
between two ferromagnetic domains. In this case, the
ferromagnetic order is preserved within one domain, and the
translational symmetry is spontaneously broken. Domain structures are
believed to appear universally in ferromagnetic systems with an energy
gap and with a finite correlation length.
On the other hand, spiral structures appear in special situations. Such
structures are extended over the entire space, and sometimes destroy
ferromagnetic order. In this letter, we construct a set of exact
ground states in a class of Hubbard-like models with coexisting domain
wall and spiral structures, which have never been found in other models. Some
remarkable results for ferromagnetic ground states have been obtained in
a class of Hubbard models recently. Mielke and Tasaki showed
independently that the ground state gives saturated ferromagnetism in a
many-electron model on a lattice with special properties, which is
called the flat-band Hubbard model \cite{Mielke, T0}. Tasaki also proved the
stability of the ferromagnetism against a perturbation which bends the
electron band \cite{T}. Tanaka and Ueda have shown this stability for a
two-dimensional model in Mielke's class \cite{TU}.
If domain wall structures are universal, then a domain wall solution
within this framework should give us some important physical
insight into theories of many-electron systems. 
Here, we construct such a solution. We deform a flat-band
Hubbard model by introducing a complex anisotropy parameter $q$. The SU(2) spin
rotation symmetry in the original flat-band model is reduced to U(1) in
our deformed model. This anisotropy $|q| \neq 1$ leads to a localized
domain wall with a finite width and a complex $q$ leads to an extended
spiral state.
We study the stability of the domain wall ground states
against a band-bending perturbation using a
variational argument. We prove that for $|q| \neq 1$, the
energy expectation value of a state with a domain wall centered near the
origin becomes lower than the eigenvalue of the saturated ferromagnetic
eigenstate, unlike the SU(2) invariant model.
We discuss similarities of the wall solution 
and differences between the domain wall solution
of our model and domain walls in pure quantum spin systems. Alcaraz,
Salinas and Wreszinski constructed a set of exact ground states with two
domains in the XXZ model with a critical boundary field in arbitrary
dimensions for an arbitrary spin \cite{ASW}.
They showed that the degeneracy of the ground states corresponding to
the location of a domain wall center is identical to that of
the ground states in the SU(2) invariant model. In their solution, the
domain wall is localized at an arbitrary surface with a finite width
depending on the Ising anisotropy parameter $(q+q^{-1})/2 > 1$. We
will see that the domain wall ground state in our electron model has the
same degeneracy as that in the XXZ model and the same localization
property in a certain parameter regime. On the other hand, for complex
$q$, 
our model differs from the XXZ quantum spin model which has no spiral
ground state.
Finally, we study from a microscopic viewpoint the scattering of a
conduction electron by the domain wall and spiral spins.

\begin{figure}[htbp]
 \begin{center}
 \setlength{\unitlength}{0.5cm}%
 \begin{picture}(8.5,7)(-4,-2)
 \put(0,1.53){\makebox(0,0){\bf\Huge$\uparrow$}}
 \put(0,3.46){\circle*{0.2}}
 \put(-2,0){\circle*{0.2}}
 \put(2,0){\circle*{0.2}}
 \put(-2.5,0){\line(1,0){5}}
 \qbezier[200](-2,0)(-1,1.73)(0,3.46)
 \qbezier[200](2,0)(1,1.73)(0,3.46)
 \qbezier[50](-2,0)(-2.25,0.435)(-2.5,0.87)
 \qbezier[50](2,0)(2.25,0.435)(2.5,0.87)
 \qbezier[20](-0.500,2.598)(-0.565,2.357)(-0.629,2.115)
 \qbezier[20](-0.500,2.598)(-0.677,2.421)(-0.853,2.245)
 \qbezier[20](0.500,2.598)(0.565,2.357)(0.629,2.115)
 \qbezier[20](0.500,2.598)(0.677,2.421)(0.853,2.245)
 \qbezier[20](-1.5,0)(-1.259,0.065)(-1.017,0.129)
 \qbezier[20](-1.5,0)(-1.259,-0.065)(-1.017,-0.129)
 \put(0,3.76){\circle{0.6}}
 \put(-2,-0.30){\circle{0.6}}
 \put(2,-0.30){\circle{0.6}}
 \put(0.3,3.46){\makebox(0,0)[l]{$x \in \mbox{even}$}}
 \put(-2.4,0.1){\makebox(0,0)[rb]{$x-1$}}
 \put(2.4,0.1){\makebox(0,0)[lb]{$x+1$}}
 \put(-0.8,2){\makebox(0,0)[r]
 {\small$t\lambda q^{-1/4}$}}
 \put(1,2){\makebox(0,0)[l]{\small$t\lambda q^{1/4}$}}
 \put(0,0.1){\makebox(0,0)[b]{\small$t e^{{\rm i} \theta /2}$}}
 \put(-2,-0.6){\makebox(0,0)[t]
 {\tiny $t(|q|^{\frac{1}{2}}+|q|^{-\frac{1}{2}})$}}
 \put(2,-0.6){\makebox(0,0)[t]
 {\tiny $t(|q|^{\frac{1}{2}}+|q|^{-\frac{1}{2}})$}}
 \put(0,4.16){\makebox(0,0)[b]
 {\scriptsize $t\lambda^2$}}
 \put(0,-1.5){\makebox(0,0)[t]{(a)}}
 \end{picture}
 \begin{picture}(8.5,7)(-4.5,-2)
 \put(0,1.53){\makebox(0,0){\bf\Huge$\downarrow$}}
 \put(0,3.46){\circle*{0.2}}
 \put(-2,0){\circle*{0.2}}
 \put(2,0){\circle*{0.2}}
 \put(-2.5,0){\line(1,0){5}}
 \qbezier[200](-2,0)(-1,1.73)(0,3.46)
 \qbezier[200](2,0)(1,1.73)(0,3.46)
 \qbezier[50](-2,0)(-2.25,0.435)(-2.5,0.87)
 \qbezier[50](2,0)(2.25,0.435)(2.5,0.87)
 \qbezier[20](-0.500,2.598)(-0.565,2.357)(-0.629,2.115)
 \qbezier[20](-0.500,2.598)(-0.677,2.421)(-0.853,2.245)
 \qbezier[20](0.500,2.598)(0.565,2.357)(0.629,2.115)
 \qbezier[20](0.500,2.598)(0.677,2.421)(0.853,2.245)
 \qbezier[20](-1.5,0)(-1.259,0.065)(-1.017,0.129)
 \qbezier[20](-1.5,0)(-1.259,-0.065)(-1.017,-0.129)
 \put(0,3.76){\circle{0.6}}
 \put(-2,-0.30){\circle{0.6}}
 \put(2,-0.30){\circle{0.6}}
 \put(0.3,3.46){\makebox(0,0)[l]{$x \in \mbox{even}$}}
 \put(-2.4,0.1){\makebox(0,0)[rb]{$x-1$}}
 \put(2.4,0.1){\makebox(0,0)[lb]{$x+1$}}
 \put(-0.8,2){\makebox(0,0)[r]
 {\small$t\lambda q^{1/4}$}}
 \put(1,2){\makebox(0,0)[l]{\small$t\lambda q^{-1/4}$}}
 \put(0,0.1){\makebox(0,0)[b]{\small$t e^{-{\rm i} \theta /2}$}}
 \put(-2,-0.6){\makebox(0,0)[t]
 {\tiny $t(|q|^{\frac{1}{2}}+|q|^{-\frac{1}{2}})$}}
 \put(2,-0.6){\makebox(0,0)[t]
 {\tiny $t(|q|^{\frac{1}{2}}+|q|^{-\frac{1}{2}})$}}
 \put(0,4.16){\makebox(0,0)[b]
 {\scriptsize $t\lambda^2$}}
 \put(0,-1.5){\makebox(0,0)[t]{(b)}}
 \end{picture}
 \end{center}
 \caption{Arrows and circles depict hopping amplitudes and on-site potentials,
respectively, for electrons with up(a) and down(b) spins on a unit cell.
 For complex $q=|q| e^{{\rm i} \theta}$, the hopping amplitude
opposite to the arrow is the complex conjugate of that along the arrow.}
 \label{fig1}
\end{figure}
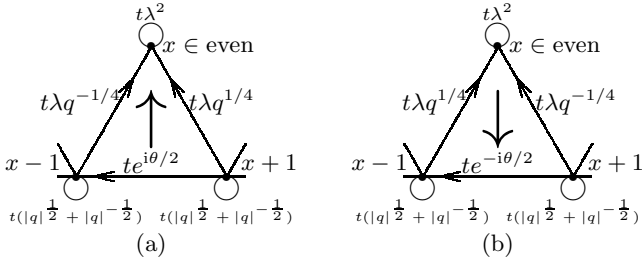

Here, we consider a one-dimensional lattice with a site index
$x= -L-1, \cdots, L + 1$, where $x$ is an integer and $L$ is an odd
integer. Electron operators on the lattice satisfy the anticommutation
relation
\[
 \{c_{x \sigma}, c_{y \tau} ^{\dag} \}
 = \delta_{\sigma \tau} \delta_{x y}, \ \ \ \
 \sigma, \tau = \uparrow, \downarrow.
\]
We define the following electron operators for even site $x$ with a
complex number $q$ and a real number $\lambda > 0$
\begin{align}
 &
 a_{x-1 \uparrow} \equiv -(q^{1/4})^{\ast} c_{x-2 \uparrow } +
 \lambda c_{x-1 \uparrow}-
 (q^{-1/4})^{\ast} c_{x \uparrow}, \nonumber \\
 &
 a_{x-1 \downarrow } \equiv -(q^{-1/4})^{\ast} c_{x-2 \downarrow} +
 \lambda c_{x-1 \downarrow }-(q^{1/4})^{\ast} c_{x \downarrow}, \nonumber \\
 &
 d_{x \uparrow} \equiv q^{-1/4} c_{x-1 \uparrow} + \lambda c_{x \uparrow}+
 q^{1/4}c_{x+1 \uparrow}, \nonumber \\
 &
 d_{x \downarrow} \equiv q^{1/4}c_{x-1 \downarrow} +
 \lambda c_{x \downarrow}+q^{-1/4}c_{x+1 \downarrow}, \\
 &
 n_{x \uparrow} \equiv c_{x \uparrow} ^{\dag} c_{x \uparrow}, \ \ \
 n_{x \downarrow} \equiv c_{x \downarrow} ^{\dag} c_{x \downarrow}, \ \ \
 n_x \equiv n_{x \uparrow}+n_{x \downarrow}.
\end{align}
The operators $a_{x \sigma}$ and $d_{y \tau} ^{\dag}$ are always anticommuting
\begin{equation}
 \{a_{x \sigma}, d_{y \tau} ^{\dag} \}
 =0, \ \ \ \ x = {\rm odd}, \ y = {\rm even}, \ \
 \sigma, \tau = \uparrow, \downarrow. \label{AC}
\end{equation}
The Hamiltonian is written in terms of the above operators as
\begin{align}
 & H = H_{\rm hop}+ H_{\rm int} \\
 & H_{\rm hop} = t \sum_{x={\rm even}} (
 d_{x \uparrow } ^{\dag} d_{x \uparrow } +
 d_{x \downarrow} ^{\dag} d_{x \downarrow})
 =\sum_{x,y,\sigma} t^\sigma _{x y} c_{x \sigma}^\dag c_{y \sigma} \\
 & H_{\rm int} =
 U \sum_x n_{x \uparrow } n_{x \downarrow},
\end{align}
with a repulsive coupling constant $U > 0$ and a hopping parameter
$t > 0$. The anticommutativity and the absence of $a_{x \sigma}$
in $H_{\rm hop}$ give a band flatness of the electrons
created by $a_{x \sigma} ^\dag$.
Here we use an open-boundary condition under which there are no degrees of
freedom at the edge sites $x=-L-2$ and $x= -L+2$, namely
$c_{-L-2 \sigma}=0=c_{L+2 \sigma}$. The hopping amplitudes on a unit
cell defined by $H_{\rm hop}$ are depicted in Fig. \ref{fig1}.


Let us discuss the symmetry of this Hamiltonian. The first important
symmetry is U(1) symmetry. The Hamiltonian commutes with the total
magnetization $S^{(3)}$, where
we define $ S^{(j)} \equiv \sum_{x} S_x^{(j)} $ and
$S_x^{(j)} \equiv \frac{1}{2} \sum_{\alpha,\beta=\uparrow, \downarrow}
 c_{x \alpha}^{\dag} \sigma_{\alpha \beta}^{(j)} c_{x \beta}$,
where $\sigma^{(j)}$ $(j = 1, 2, 3)$ are the Pauli matrices.
In the limit $q \rightarrow 1$, this symmetry is
enhanced to the SU(2) symmetry. In this limit, this model becomes the
original flat-band Hubbard model given by Tasaki \cite{T0,T}. The
second important symmetry is 
defined by a product of a parity $P$ and a spin rotation
\begin{align}
 \Pi=\Pi^{-1} = & P \exp \left( {\rm i} \pi S^{(1)} \right) \label{pi} \\
 \Pi c_{x \sigma} \Pi = c_{-x \ -\sigma} ,&
 \Pi c_{x \sigma}^{\dag} \Pi = c_{-x ~-\sigma}^{\dag}, \nonumber
\end{align}
Note the following transformation of the total magnetization
$\Pi S^{(3)} \Pi =-S^{(3)}.$ An energy
eigenstate with the total magnetization $M$ is transformed by $\Pi$ into
another eigenstate with the total magnetization $-M$, which belongs to
the same energy eigenvalue.


Now we construct ground states in the quarter-filled case,
where the uniqueness of the ground state can be shown.
Since each term in the Hamiltonian is positive semi-definite
$H_{\rm hop} \geq 0$ and $H_{\rm int} \geq 0$, an eigenstate with a zero-energy 
eigenvalue is a ground state. An arbitrary state created only by
$a_{x \uparrow}^{\dag}$ is a zero-energy states because of its
anticommutativity (\ref{AC}) with $d_{x \uparrow}$ and no double
occupancy, and thus this polarized state
\begin{equation}
 |{\rm all ~up} \rangle = \prod_{x= {\rm odd}} a_{x \uparrow } ^\dag
 | {\rm vac} \rangle,
 \label{AU}
\end{equation}
is a ground state. To create ground states
with other magnetization, first we should take into account
a condition of no double occupancy at odd sites.
A state created by $a_{x \sigma} ^\dag$ under this condition
is written in the following summation over all spin
configurations with an arbitrary fixed magnetization
$\sum_x \sigma_x =M$
\[
 |M \rangle = \sum_{ \sigma_{-L} + \cdots+ \sigma_L = M }
 \psi(\sigma_{-L}, \cdots, \sigma_L) \prod_{x={\rm odd}}
 a_{x \sigma_x} ^\dag | {\rm vac} \rangle.
\]
The condition of no double occupancy on the state $| M \rangle$ at
even sites yields
\[
 \psi(\cdots, \uparrow, \downarrow, \cdots)=
 q~ \psi(\cdots, \downarrow, \uparrow, \cdots).
\]
This relation implies the uniqueness of the ground state with a fixed
total magnetization, since two arbitrary spin configurations can be
related by the successive exchanges of two nearest neighbor spins.
Therefore the degeneracy of those ground states is exactly the same as
that in the SU(2) symmetric model, as in the domain wall solution in
quantum spin systems \cite{ASW}.

To explore the nature of the ground state, we write it in a
more explicit way. 
The following superposition
over the states with different magnetizations can be a ground state
\begin{equation}
 | z \rangle \equiv \sum_{M} z^{L-M}|M>
 =\prod_{x= {\rm odd}}(a_{x \uparrow}^{\dag}+ z
 q^{\frac{x}{2}}a_{x \downarrow}^{\dag})
 |{\rm vac} \rangle, \label{z},
\end{equation}
where $z$ is an arbitrary complex number. This state can be regarded as
a generating function for a ground state with an arbitrary
magnetization.
\begin{figure}[b]
 \begin{center}
 \rotatebox{90}{$\langle S_{x}^{(3)} \rangle$}%
 \begin{minipage}[c]{80mm}
 \includegraphics[width=80mm]{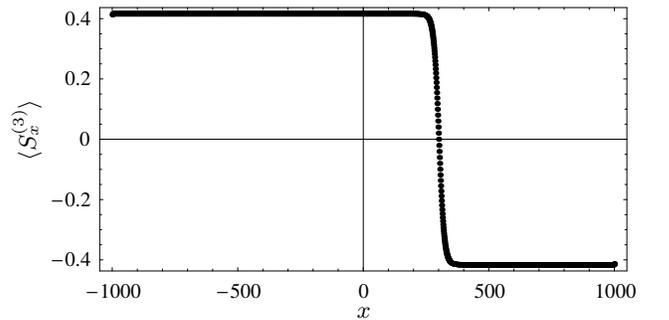}
 \end{minipage}

 \hspace*{3em}$x$
 \end{center}
 \vspace*{-1.5em}
 \caption{Spin one point function on odd sites $x$ evaluated
 from the numerical solution of the recursion relation for
 $L = 1000$, $q = 1.05$, $\lambda = 3$ and $z = 1.05^{-300}$.}
 \label{wall}
\end{figure}
The expectation values of the spin operators at site $x$ in this ground
state are
\begin{align}
 &\Bigl\langle S_{x} ^{(1)} \Bigr\rangle= \frac{\langle n_{x} \rangle}{2}
 \frac{z q^{\frac{x}{2}} +(z q^{\frac{x}{2}})^{\ast}}{1+|z^2 q^{x}|}
 \nonumber \\
 &\Bigl\langle S_{x} ^{(2)} \Bigr\rangle
 = \frac{\langle n_{x} \rangle}{2 {\rm i}}
 \frac{z q^{\frac{x}{2}} -(z q^{\frac{x}{2}})^{\ast}}{1+|z^2 q^{x}|}
 \nonumber \\
 &\Bigl\langle S_{x} ^{(3)} \Bigr\rangle
 =
 \frac{\langle n_{x} \rangle}{2}\frac{1-|z^2 q^{x}|}{1+|z^2 q^{x}|},
\label{spin}
\end{align}
where the expectation value of an operator ${\cal O}$ is defined by
$\langle {\cal O} \rangle \equiv \langle z | {\cal O} | z \rangle/
\langle z | z \rangle$. As discussed in the XXZ models
\cite{KN1,KN2,M,KNS},
the two domains are distinguished by the
sign of the local order parameter $\langle S_{x} ^{(3)}
\rangle$. The domain wall center is defined by the zeros
of $\langle S_{x} ^{(3)} \rangle $ which is located
at $x= -2 \log_{|q|}|z|$. The function
$\frac{1}{2} \langle n_{x}\rangle -|\langle S_{x} ^{(3)} \rangle |$
decays exponentially from the center, if $\langle n_{x}\rangle \neq 0$.
 The domain wall
width $1 / \log |q|$ is defined by its decay length.
A profile of the domain wall in a typical example
is depicted in Fig.\ref{wall}.
For the complex $q=|q|e^{{\rm i} \theta}$,
one can see the spiral structure with a pitch angle $\theta$.
The vector $\langle \vec S_x \rangle$
is rotated with angle $\theta x$
around the third spin axis depending on the site $x$.
Note that this spiral structure of the ground state does not exist in the
XXZ model. Though the complex anisotropy parameter
$q=e^{{\rm i} \theta}$ is possible in the XXZ Hamiltonian, the spiral state
is no longer the ground state and the corresponding model
is described in the Tomonaga-Luttinger liquid without ferromagnetic order.
The translational symmetry
in the infinite volume limit is
broken by the domain wall or the spiral structure for finite $\log |z|$.
Both symmetries generated by $S^{(3)}$ and $\Pi$ are
broken spontaneously as well.

Now, we evaluate the function
$\langle n_x \rangle$ to see the practical form of
$\langle \vec S_x \rangle$ given by eq. (\ref{spin}).
First we consider a simple case of the large $\lambda$ limit.
This limit
implies that there is no overlapping of the electrons at each even site, thus
\begin{equation}
 \langle n_x \rangle = 
 1-O(\lambda^{-2}),~ x ={\rm odd}, ~~
 \langle n_x \rangle = 
 O(\lambda^{-2}),~ x = {\rm even}.
 \label{loc}
\end{equation}
The localization property of the ground state in this case is the same as in the XXZ model.
To treat a general case, we introduce the following normalization function
\begin{equation}
 A(x_0, x, \zeta) \equiv \|
 \prod_{y=x_0} ^x (a_{y \uparrow}^{\dag}+ |q|^{-\frac{\zeta}{2}}
 q^{\frac{y}{2}}a_{y \downarrow}^{\dag})
 |{\rm vac} \rangle \|^2,
\end{equation}
where we employ $z$
as a real number and
use a parameterization $z^2 = |q|^{-\zeta}$ for $|q| > 1$.
Note $A(-L, L, \zeta)= \langle z | z \rangle$.
This normalization function obeys the recursion relation
\begin{align}
 A(x_0, x, \zeta) = &
 \epsilon ~(1 + |q^{x-\zeta}|)A(x_0,x-2, \zeta) \nonumber \\
 & - (1+|q^{x-\zeta-1}|)^2 A(x_0, x-4, \zeta),
\label{A}
\end{align}
where $\epsilon \equiv \lambda^2+ |q|^{1/2}+|q|^{-1/2}$.
We can extract the main $x$ dependent part out of
$A(x_0, x, \zeta)$ as follows
\begin{equation}
 A(x_0, x, \zeta) = B(x_0, x, \zeta) r^{\frac{x-x_0+2}{2}}
 \prod_{y= x_0} ^x (1 + |q^{y-\zeta}|),
\end{equation}
where
$r \equiv (\epsilon + \sqrt{\epsilon ^2 -4})/2$.
The convergence of the quotient $B(-L, x, \zeta)$ in the
infinite volume limit is proved rigorously
on the basis of the recursion relation for $B(-L, x, \zeta)$
derived from the recursion relation (\ref{A}) \cite{HI}.
The one point function is represented in terms of the normalization function
\[
 \langle n_x \rangle =
 \frac{\lambda^{2}}{r}
 \frac{B(-L, x-2, \zeta) B(x+2, L, \zeta)}{B(-L,L, \zeta)},
\]
for odd $x$. The expression for even $x$ is also easily obtained 
as a closed but complicated form.
We have bounds
$l \leq \langle n_x \rangle \leq u $
with some positive constants $l$ and $u$ less than 1.
Therefore, these bounds and
eq. (\ref{spin}) guarantee the unique domain wall center at $x=\zeta$.
The numerical solution of the recursion relation for $B(-L, x, \zeta)$ is useful
for seeing the profile of the domain wall practically.
The spin one point function for
$\langle S_x^{(3)}\rangle$ is depicted in Fig.\ref{wall}.
Now, we study the stability of the domain wall and the spiral structure of
the ground state against a band bending perturbation in a variational argument. 
To bend the lower flat band, we consider the following
perturbation
\begin{equation}
 H_{\rm hop} ' = -s \sum_{x,\sigma} a_{x \sigma} ^{\dag} a_{x \sigma},
\end{equation}
with a positive small constant $s$. As has been rigorously proved for $q=1$ by
Tasaki \cite{T}, the all-spin-up state (\ref{AU}) is preserved as a ground
state for sufficiently small $s/t$ and $s/U$. Here, we consider the
problem also in this limit.
The all-spin-up state (\ref{AU}) is an eigenstate of the Hamiltonian
$H=H_{\rm hop}+H_{\rm int}+H_{\rm hop}'$ still
\[
 H |{\rm all ~up} \rangle = E({\rm all ~up})~
 |{\rm all ~up} \rangle,
\]
where
$E({\rm all ~up}) = - (L+1) s \epsilon$. If
$|q| \neq 1$, other domain wall states (\ref{z}) are not eigenstates,
and the energy expectation values are lower than this energy eigenvalue
$E({\rm all ~up})$. First, we consider the case of $|q| \neq 1$.
In the large $t$ and large $U$ limit,
the energy expectation value is written in terms of the
normalization functions
\begin{equation}
 E(\zeta) \equiv \frac{\langle z | H | z \rangle}{\langle z | z \rangle}
= E({\rm all ~up}) - s \sum_{x=-L} ^{L}
 g(x,\zeta),
\end{equation}
where $x$ is summed over odd integers and
\begin{align}
 &g(x, \zeta) \equiv
 \frac{(|q|^\frac{1}{2}-|q|^{-\frac{1}{2}})^2 B(-L,x-4, \zeta)}{r^2
 (|q^\frac{x-\zeta-2}{2}|+
 |q^{-\frac{x-\zeta-2}{2}}|)B(-L,L,\zeta)} \\
 &\times
 \left[
 \frac{2 \epsilon B(x+2, L, \zeta)}
 {|q^\frac{x-\zeta}{2}|+|q^{-\frac{x-\zeta}{2}}|}
 - \frac{(|q|^{\frac{1}{2}}+|q|^{-\frac{1}{2}})^2 B(x+4,L,\zeta)
 }{r
 (|q^{\frac{x-\zeta+2}{2}}|+|q^{-\frac{x-\zeta+2}{2}}|)}
 \right] .\nonumber
 \label{g}
\end{align}
Let us compare the energy expectation values of two solutions with 
different domain wall centers, namely,$\zeta$ and $\zeta+2$.
The analysis of the recursion relation in \cite{HI} enables us to
estimate $B(-L, x, \zeta)$, then
the energy difference of two nearest neighbor domain
walls with a finite distance $\zeta = O(1)$ from the origin is
\[
|E(\zeta +2) - E(\zeta)| = O(L |q|^{-L} s).
\]
The energy cost by the finite shift of the domain wall center might vanish in
the infinite volume limit. If this is true, the domain wall
solution with an arbitrary center is stable against the perturbation
$H_{\rm hop} '$. The degeneracy corresponding to the location of the
domain wall center may be preserved in the infinite volume limit. On
the other hand the energy difference between the all-spin-up state and
the state with a domain wall at $\zeta=-L+1$ is
\[
 E({\rm all~up})-E(-L+1)=O(s).
\]
The energy eigenvalue of the all-spin-up state is larger than the domain
wall state with order $1$. Since the energy of the true ground state is
lower than the expectation value of the trial state, the all-spin-up
state is no longer a ground state unlike the
SU(2) invariant model. Therefore,
in this variational argument, we conjecture that the localized domain
wall structure is stable against this band-bending perturbation.
On the other hand, for $|q|=1$, all the states $|z \rangle$ are
still exact eigenstates and are still degenerate to the eigenvalue
$E({\rm all ~up})$. Therefore, the stability of the ground states can be
proved in the same way as that given by Tasaki \cite{T}.


Next, we consider the behavior of one electron added to
the quarter filled ground state
$| z \rangle$ with a domain wall or a spiral structure,
\begin{equation}
 |\psi \rangle \equiv
 \sum_{x, \sigma} \psi_{x \sigma} c_{x \sigma} ^\dag | z \rangle
\end{equation}
According to a variational principle, we find the trial wave function
$\psi_{x \sigma}$ to optimize the following variational functional
\begin{align}
 \langle & \psi |H | \psi \rangle
 = 
 \sum_{x,y} t^\sigma _{x y} \psi_{x \sigma}^{\ast} \psi_{y \sigma}
 \nonumber \\
 &
 + U \sum_{x}
 \Biggl(
 \frac{|z^2 q^{x}|}{1+|z^2 q^{x}|} \psi_{x \uparrow}^{\ast} \psi_{x \uparrow}
 +\frac{1}{1+|z^2 q^{x}|} \psi_{x \downarrow}^{\ast} \psi_{x \downarrow}
 \nonumber \\
 &
 -\frac{z q^{\frac{x}{2}}}{1+|z^2 q^{x}|} \psi_{x \downarrow}^{\ast}
 \psi_{x \uparrow}
 -\frac{(z q^{\frac{x}{2}})^{\ast}}{1+|z^2 q^{x}|}
 \psi_{x \uparrow}^{\ast} \psi_{x \downarrow}
 \Biggr) \langle n_x \rangle.
\end{align}
In our model, the scattering of conduction electrons by the
domain wall or spiral spins of the lower band has been given already
as a well-defined problem.
By optimizing this effective energy function under a
normalization constraint $\langle \psi |\psi \rangle = 1$, we obtain a
two-component Schr\"odinger equation. At large $\lambda$ or for 
 $|q|=1$, the $\langle n_x \rangle$ becomes periodic for the lattice unit
cells. In particular, at large $\lambda$, eq. (\ref{loc})
implies that a localized spin sits at an odd site
and the conduction electron hops mainly on even sites
at large $U$. 
The normalization constraint becomes local in this limit and the hopping
term gives a conventional kinetic term in continuum approximation, and
the equation for real $q > 1$ becomes that for a conduction electron on
the Kondo lattice coupled to the XXZ model obtained by Yamanaka and Koma
\cite{YK}. The model on the Kondo lattice is realized under special
conditions in our model.


In this letter, we construct a set of exact ground states with
a ferromagnetic domain wall and a spiral
structure in a deformed flat-band Hubbard model.
Some results obtained here can be extended to
higher dimensional models, as discussed in the original
flat-band Hubbard model \cite{T0}.
After the construction of the domain wall
ground state in the XXZ model \cite{ASW},
interesting properties of
excitations were discovered \cite{KN1,KN2,M,KNS}.
In particular, singular low-lying excitations
localized near the domain wall
were discussed in arbitrary dimensions \cite{M,KNS}.
We can expect similar properties
of excitations above the domain wall in our electron model as well.

\begin{acknowledgments}
The authors are grateful to C. Nayak and A. Tanaka for carefully reading
the manuscript and kind suggestions. They would like to thank T. Koma,
B. Nachtergaele, H. Tasaki and M. Yamanaka for helpful comments.
\end{acknowledgments}


\begin{thebibliography}{99}
 \bibitem{Mielke} A. Mielke:
 J. Phys. A {\bf 24} (1991) L73.
 \bibitem{T0} H. Tasaki:
 Phys. Rev. Lett. {\bf 69} (1992) 1608.
 \bibitem{T} H. Tasaki:
 Phys. Rev, Lett. {\bf 75} (1995) 4678.
 \bibitem{TU} A. Tanaka and H. Ueda:
 Phys. Rev. Lett. {\bf 90} (2003) 067204.
 \bibitem{ASW} F. C. Alcaraz, S. R. Salinas and W. F. Wreszinski:
 Phys. Rev. Lett. {\bf 75} (1995) 930.
 \bibitem{KN1} T. Koma and B. Nachtergaele:
 Lett. Math. Phys. {\bf 40} (1997) 1.
 \bibitem{KN2} T. Koma and B. Nachtergaele:
 Adv. Theor. Math. Phys. {\bf 2} (1998) 533.
 \bibitem{M} T. Matsui:
 Lett. Math. Phys. {\bf 37} (1996) 397.
 \bibitem{KNS} T. Koma, B. Nachtergaele and S. Starr:
 Preprint, math-ph/0110017.
 \bibitem{HI} M. Homma and C. Itoi:
 in preparation.
 \bibitem{YK} M. Yamanaka and T. Koma:
 J. Magn. Soc. Jpn. {\bf 23} (1999) 141.
\end{thebibliography}
\end{document}